# LEDA BEAM OPERATIONS MILESTONE AND OBSERVED BEAM TRANSMISSION CHARACTERISTICS*

L. J. Rybarcyk, J. D. Schneider, H. V. Smith, and L. M. Young, M. E. Schulze[a],
Los Alamos National Laboratory, Los Alamos, NM 87544, USA.

*Abstract*

Recently, the Low-Energy Demonstration Accelerator (LEDA) portion of the Accelerator Production of Tritium (APT) project reached its 100-mA, 8-hr CW beam operation milestone. LEDA consists of a 75-keV proton injector, 6.7-MeV, 350-MHz CW radio-frequency quadrupole (RFQ) with associated high-power and low-level rf systems, a short high-energy beam transport (HEBT) and high-power (670-kW CW) beam dump. During the commissioning phase it was discovered that the RFQ field level must to be approximately 5-10% higher than design in order to accelerate the full 100-mA beam with low losses. Measurements of a low-duty-factor, 100-mA beam show the beam transmission is unexpectedly low for RFQ field levels between ~90 and 105% of design. This paper will describe some aspects of LEDA operations critical to achieving the above milestone. Measurement and simulation results for reduced RFQ beam transmission near design operating conditions are also presented.

## 1 INTRODUCTION

The LEDA RFQ is an 8-m-long linac that delivers a 6.7-MeV, 100-mA CW proton beam. The RFQ and its various ancillary systems (Fig. 1) are described in detail elsewhere [1-6]. We recently completed the beam-commissioning phase where we accomplished our goal of 8-hr, 100-mA, CW beam operation. During the coarse of commissioning the RFQ, we observed an overall reduction in beam transmission for peak currents >70 mA and RFQ field levels between ~90 and 105% of design. This paper presents results obtained during the 100-mA CW beam-commissioning phase. It also summarizes numerous measurements and simulations performed in an attempt to understand the aforementioned loss of transmission.

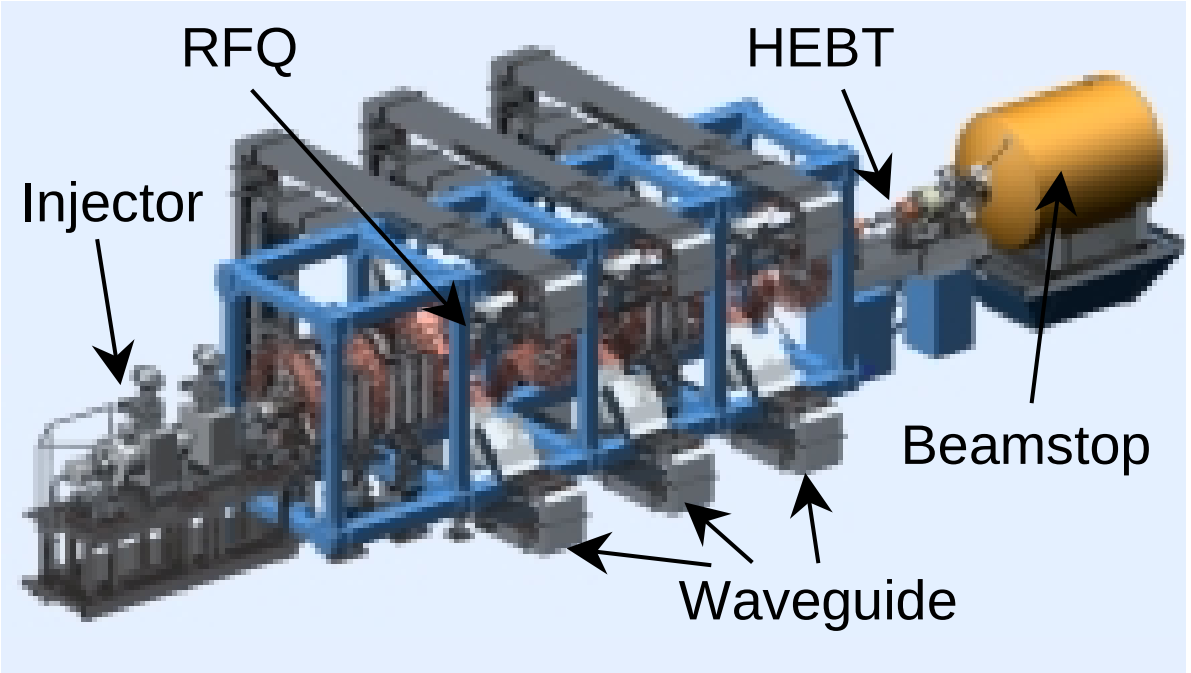

Figure 1: LEDA configuration for RFQ commissioning. Center and lower-upstream waveguides no longer used.

## 2 LEDA PERFORMANCE

From mid-Nov ’99 through early Apr ’00, LEDA was operated with beam currents in excess of 90 mA and duty factors ≥99.7%. During this time, while operating at duty factors ≥99.7%, LEDA accumulated 9.0 hr of ≥99.7 mA, 20.7 hr of ≥99 mA and 111 hr of ≥90 mA beam. The beam-current monitors were sampled at 30-sec intervals. In the analysis, a run was defined as a contiguous set of samples with the beam current ≥5 mA and the duty factor ≥99.7%. To aid in monitoring beam transmission through the RFQ, the duty factor was reduced from CW to 99.7% to allow accurate measurement of the RFQ input and output beam using the AC current monitors at the entrance and exit of the RFQ. These data are included in the CW analysis. The longest run at greater then 90 mA was 118 min at 99.3 mA. We accumulated a total of 694 runs with an average duration of 9.6 min each. A histogram of run duration statistics is shown in Fig. 2.

During the commissioning phase, the following aspects of either LEDA hardware configuration or operations were critical to achieving the 100-mA milestone:

- LEBT beam properly matched to the RFQ. Space-charge effects were overcome through addition of an electron-trap at the RFQ entrance and by reduction of final LEBT solenoid to RFQ distance [1].
- RFQ field quality. Monitored at 64 locations along structure and optimized through adjustment of cooling flows on the four 2-m RFQ segments.
- Resonance Control Cooling System performance. PID control parameters were adjusted for fast transient response during high-power beam operation.
- Low operating pressure in RFQ. Pressure ~$1.x10^{-7}$ Torr for stable operation.

* Work supported by US Department of Energy

[a] General Atomics, Los Alamos, NM 87544, USA.

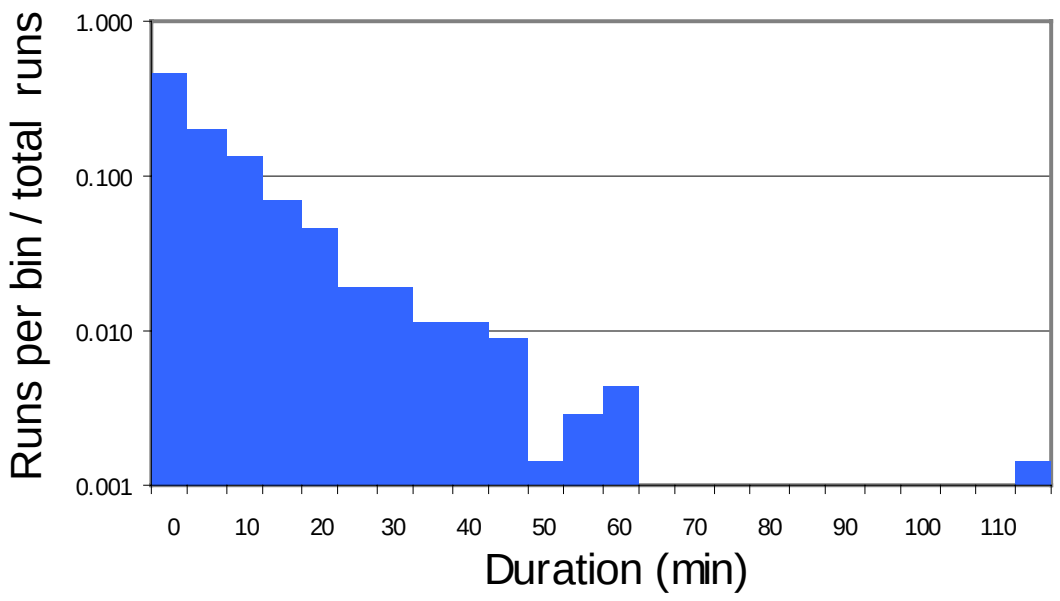

Figure 2: Histogram showing likelihood of success for a given duration run. Data set includes all CW runs of ≥90 mA. Data binned in 5 min intervals.

- HEBT tuned for 100-mA beam. Facilitated using “notched” RF in synch with injector pulse to produce short, high-current pulses for tuning.
- High (>90%, design:~93%) beam transmission. Below ~90% the losses were too great to sustain stable operation.
- RFQ field levels at 5-10% above design.

## 3 RFQ BEAM TRANSMISSION

### *3.1 Initial Observation*

Early on in the commissioning phase, an RFQ transmission curve first revealed a discrepancy between actual and expected beam transmission at high peak currents. Subsequently a series of measurements were made to examine the transmission for various peak current beams from 70-100 mA. The measurement results along with a PARMTEQM [7] prediction for the nominal 100 mA beam are shown in Fig. 3. All transmission measurements were performed with low duty factor beam to limit the total beam loss. A substantial peak-current-dependent reduction in transmission was observed. This reduction in transmission for the 100mA beam ultimately dictated a higher operating field level for the RFQ, i.e. field levels ~5-10% above design.

### *3.2 Simulations*

In an attempt to understand the loss of transmission at higher beam currents, numerous PARMTEQM calculations were performed to investigate the effects of varying degrees of RFQ field tilt, beam mismatch, position and angle offsets to the beam, and beam current enhancement on RFQ beam transmission. None of the above results were able to reproduce the observed loss of transmission. Initially, simulations were performed with field tilts up to 10%. These results did not reproduce the measurements. Also, these large tilts were not consistent with our observations. (Quadrupole and dipole field distributions were derived from cavity signals sampled at 64 locations along the RFQ.) Introducing mismatched beam into the RFQ reduced the overall transmission but nothing more. The code was then modified to allow for small displacements to be applied to the particles transverse coordinates at a given cell. This was an attempt to approximate small misalignments between segments of the RFQ and small dipole field contributions from the RFQ. These results did not reproduce the transmission curves. The code was further modified to allow the beam current in a bunch to be enhanced. This was an attempt to mimic background charges that might possibly become trapped within the RFQ acceleration channel. In all the above studies, the character of the predicted transmission curves was basically unchanged. The precipitous drop in transmission has not been reproduced by any of these calculations.

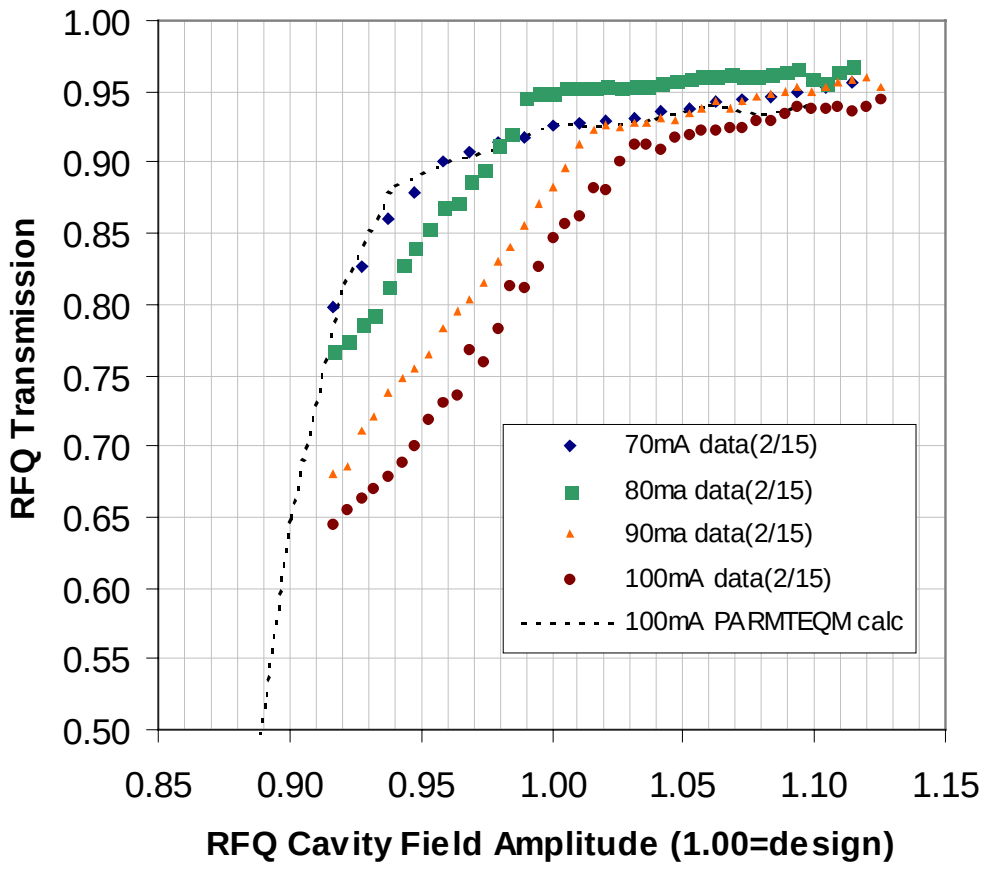

Figure 3: LEDA RFQ transmission data. Measurements performed at 70-100 mA peak current. PARMTEQM prediction for nominal 100-mA beam also shown.

### *3.3 Additional Observations*

Further measurements revealed several interesting features. Time dependence in the loss of transmission was seen while making measurements using short beam pulses at reduced RFQ field strengths. We observed a step-change reduction in the beam current out of the RFQ as shown in Fig. 4. The leading portion of the pulse exhibits transmission characteristics in agreement with simulation for the nominal RFQ, the trailing edge does not. Simultaneous measurements of the cavity field amplitude sampled along the downstream portion of the RFQ revealed a small but measurable increase in the RFQ field level correlated with the decrease in beam transmission, also shown in Fig. 4. During this transition, the low-level RF system maintained a constant drive signal. The RFQ field amplitude at the end of the pulse was observed to increase exponentially towards the end of the RFQ. This would be consistent with a reduction in beam loading, i.e. increase in beam loss, which would result in net higher fields. The beam loss would also be consistent with observed high radio-activation levels at the high-energy end of the RFQ when it is operated at or below design field levels. We also observed that the temporal location of the transition depends upon the RFQ field level. As the field is reduced the start of the transition shifts towards the beginning of the pulse. Also, no difference was seen in transmission curves obtained using 90-mA beam under CW and low duty factor(~25%) RF operation.

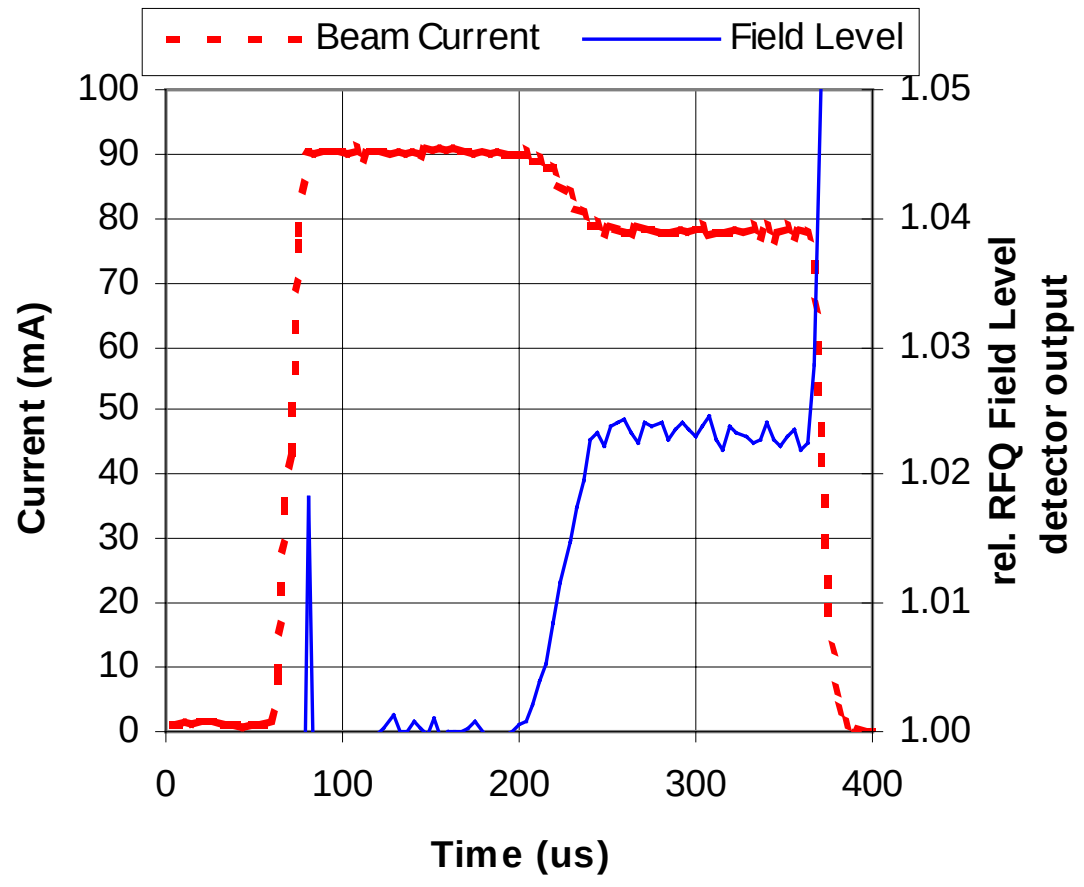

Figure 4: RFQ output current and field level versus time. RFQ nominal field level at ~97% of design.

Changes in the wire scanner profiles were also seen across a beam pulse containing the transition. The centroid and rms width of both the horizontal and vertical profiles were constant during the leading edge of the pulse. However, during the trailing edge as much as a 100% increase in the rms size of the vertical profile was seen as the field level in the RFQ was reduced from 10% above to 10% below design. Over that same field amplitude range a small change was observed for the horizontal rms size while no change was observed in either centroid.

### *3.4 One possible explanation*

The observed reduction in transmission might well be due to ions trapped within the RFQ accelerating channel. The potential well established in the RFQ is capable of trapping slow moving ions [8]. These ions would increase the effective space-charge force seen by the beam and could result in a larger overall beam that could be lost on the RFQ vanes. The source of ions, e.g. protons, might be beam collisions with either residual gas molecules or the RFQ vane tips. The observed time dependence in the output beam current would be related to the ion buildup rate. Previous simulations with PARMTEQM may have been to simplistic. A further study was performed where PARMTEQM was modified to pre-load the space-charge mesh with additional charge. Very preliminary results showed a background charge distribution could produce a larger, somewhat hollow beam. Transmission calculations have not yet been performed with this code. However, a steady state, single-bunch code like PARMTEQM might not be appropriate for modeling this time dependent phenomenon. Along this line, work has begun on developing a simple model of the RFQ using time as the independent variable. A time-based code like TOUTATIS [9] might also be more appropriate for this study. More work needs to be done in this area.

## 4 SUMMARY

The LEDA RFQ has performed well: it operated for 21 hr with RFQ output currents ≥99 mA during the recent beam-commissioning period. An unexpected reduction in high-peak current (>70mA) beam transmission was observed when the RFQ field levels were operated between ~90 and 105% of design. Further investigation revealed a time dependent character to the beam transmission and correlated effects in wire scanner profiles and RFQ field levels. Trapped ions in the RFQ channel is one possible explanation for the effect.

## ACKNOWLEDGEMENTS

The authors thank the LEDA operations and support personnel without whom this work would not have been possible.